\documentstyle[preprint,eqsecnum,aps]{revtex}
 \tightenlines
\draft

\newtheorem{theorem}{Theorem}

\newcommand{\be}{\begin{equation}}
\newcommand{\ee}{\end{equation}}
\newcommand{\beal}{\begin{eqalign}}
\newcommand{\eeal}{\end{eqalign}}
\newcommand{\bea}{\begin{eqnarray}}
\newcommand{\eea}{\end{eqnarray}}
\newcommand{\bean}{\begin{eqnarray*}}
\newcommand{\eean}{\end{eqnarray*}}
\newcommand{\ba}{\begin{array}}
\newcommand{\ea}{\end{array}}

\newcommand{\ep}{\epsilon}
\newcommand{\Th}{\Theta}
\newcommand{\Ga}{\Gamma}
\newcommand{\ga}{\gamma}

\newcommand{\La}{\Lambda}

\newcommand{\de}{\delta}

\newcommand{\pa}{\partial}

\newcommand{\no}{\nonumber}

\newcommand{\res}{\mbox{res}}

\newcommand{\op}{{\mathcal{O}}}
\newcommand{\lan}{\langle}
\newcommand{\ran}{\rangle}
\newcommand{\ti}{\tilde}

\begin{document}

\title
 { \sc On the Benney Hierarchy:\\
  free energy, string equation and quantization\/}
\author{
{\sc Jen-Hsu Chang\/}\\
  {\it Institute of Mathematics, Academia Sinica,\\
   Nankang, Taipei, Taiwan\/}\\
   E-mail: changjen@math.sinica.edu.tw\\
    and\\
 {\sc Ming-Hsien Tu\/}\\
  {\it Department of Physics, National Chung Cheng University,\\
   Minghsiung, Chiayi, Taiwan\/}\\
   E-mail: phymhtu@ccunix.ccu.edu.tw
}
\date{\today}
\maketitle
\begin{abstract}
The bi-Hamiltonian structure of the Benney hierarchy is revisited.
 We show that the compatibility  condition of the Poisson brackets provides
   the genus zero free energy of a topological field theory coupled to 2d gravity.
    We calculate the correlation functions via the Landau-Ginzburg  formulation
     and derive the string equation based on the twistor construction. Moreover,
 by using the approach of Dubrovin and Zhang, we compute the genus one correction of the Poisson
 brackets and compare them with the Oevel-Strampp's brackets of the Kaup-Broer hierarchy.
\end{abstract}

\newpage

\section{Introduction}

In the past decade, the developments of integrable systems have made many important influences
 on theoretical physics and pure mathematics. Among others, those works concerning
 the relationship to topological field theories (TFT) and string theories have been paid
  much attention in the frontier subjects (see \cite{D} for a review).
   In particular,  Witten \cite{W2} and Kontsevich \cite{K}  show that the partition
    function of 2d topological gravity is equivalent to a particular tau-function of the
     Korteweg-de Vries hierarchy characterized by
  the string equation. Now it is generally believed that 2d TFT
   coupled to 2d gravity can be formulated as integrable hierarchy of nonlinear partial
    differential equations.

    In general, 2d TFT can be classified by the solutions
     of the Witten-Dijkgraaf-E.Verlinde-H.Verlinde (WDVV) equations of
     associativity \cite{W2,DVV} in the sense that a particular solution
        of WDVV equations provides the primary free energy of some topological model.
        In fact, various classes of solutions to the WDVV equations have been obtained
         (see \cite{Kr1,Du4,Du2} and references therein),
         which turn out to be the tau-functions of dispersionless integrable
         hierarchies. Accordingly, investigating the solution space of the WDVV
           equations will deepen our understanding of 2d TFT.

 It is well-known that  Poisson structures of dispersionless integrable hierarchies have
       the form of hydrodymanic type \cite{DN1}. Due to this fact, the
    integrability (bi-Hamiltonian structure) of the WDVV equations can be formulated in
     a geometric way called    Frobenius manifolds \cite{Du2}.
      Based on this geometrical construction, the higher genus extension of the Poisson structures
       \cite{DZ} and Virasoro constraints \cite{DZ2}
      of the associated integrable hierarchies are given.

   The purpose of this paper is to study a dispersionless {\it nonstandard\/}
    Lax hierarchy from the TFT point of view, which  is a modification of the
     dispersionless Kadomtsev-Petviashvili (dKP) hierarchy \cite{KG,TT1}.
       The Lax operator we would consider has the form
\bea
 L=p^N+v^{1}p^{N-1}+v^{2}p^{N-2}+\cdots+v^N+ \frac{v^{N+1}}{p}
  \label{lax2}
   \eea
    which satisfies the hierarchy flows ($T_1=X$)
     \bea
      \frac{\pa L}{\pa T_n}=\{L^{n/N}_{ \geq 1}, L \},
       \label{dmkp}
        \eea
  where $L^n_{\geq 1}$ means the polynomial $p^n+ \cdots + (\cdots)p$, i.e.,
   we cut off the terms after $p$ of the
expansion (in $p$) of $L^n $ and the Poisson bracket  $\{,\}$ is defined by \cite{LM}
 \bea
  \{f(p,X), g(p,X) \}=\frac{\pa f}{\pa p}\frac{\pa g}{\pa X}-
  \frac{\pa f}{\pa X}\frac{\pa g}{\pa p}.
  \label{defpo}
\eea
 We remark  that the Lax operator (\ref{lax2}) can be obtained from that
of the dispersionless modified KP (dmKP)  hierarchy \cite{Li,CT1} via truncations.
 Thus the nonstandard Lax hierarchy (\ref{dmkp}) is referred to the
  constrained dmKP hierarchy \cite{CT2}.

  In this paper, for simplicity, we shall concentrate on the Lax
operator of the form
\be
 L=p+v^1+v^2p^{-1}
 \label{blax}
  \ee
which satisfies the nonstandard Lax equations
 \bea
  \frac{\pa L}{\pa T_n}=\{L^{n}_{\geq 1}, L \}
   \label{ben1}
    \eea
    (we will leave the cases for $N\geq 2$ in a subsequent publication).
     The first few flows are
 \bea
 \frac{\pa}{\pa T_2}\left(
  \ba{c} v^1 \\
   v^2
 \ea
 \right) &=& \left( \ba{c} (v^1)^2+2v^2\\
2v^1v^2 \ea \right)_X, \label{ben2}\\
 \frac{\pa}{\pa T_3} \left(
  \ba{c} v^1
\\ v^2
 \ea
  \right)
  &=& \left( \ba{c} 6v^1v^2+6(v^1)^3\\
3(v^1)^2v^2+3(v^2)^2
 \ea \right)_X,\no
  \eea
where the simplest equation ($T_2$-flow) being the Benney equation
 describes long waves in nonlinear phenomena (here we only consider the $1+1$
  dimensional reduction of the Benney system in \cite{B}).  The whole equations (\ref{ben1})
 form what we call the Benney hierarchy which has been intensively studied
 during the past two decades (see, for example, \cite{EYR,FS,GT,KM,LM,Zak}
  and references therein).\\
  {\bf Remark 1.\/} The algebraic and Hamiltonian structure associated with the kind of
  Lax operators (\ref{lax2}) have also been investigated in \cite{FS} where the Lax equations
   are defined by the bracket  $\{A,B\}=p\pa A/\pa p\pa B/\pa x-p\pa A/\pa x\pa B/\pa p$
    with respect to the decomposition $\La=\La_{\ge 0}\oplus \La_{< 0}$  for the
    pseudo-differential operator $\La$ and the relationship to the dispersionless
     Toda hierarchies (dToda) \cite{TT,TT1} is  established as well.

 Recently, the bi-Hamiltonian structures associated with the Benney hierarchy
  (\ref{ben1}) have been investigated based on the theory of classical $r$-matrices
   \cite{Li}. From that one can associate with a free energy coming from a
    2d TFT. In particular, by this free energy, we  construct
     an additional hierarchy generated by  Hamiltonians with logarithmic type
      which together with the ordinary hierarchy are identified as
 flows in genus zero TFT coupled to 2d topological gravity.
  The basic idea is that the dispersionless Lax operator (\ref{blax}) can be viewed
   as a superpotential in  Landau-Ginzburg (LG) formulation of TFT \cite{V,DVV}.
   Thus, according to the LG theory, the variables $v^1$ and $v^2$ are identified as
   the fundamental correlation functions and their dynamic flows turn out to be
  the genus zero topological recursion relations \cite{W2,DW} of the associated TFT.

 Moreover, in order to establish the string equation describing the gravitational effect,
  we construct the twistor data for the Benney hierarchy by using the Orlov operator corresponding
 to the dmKP hierarchy\cite{CT1}. We show that a remarkable feature of Benney hierarchy is the
  flows generated by the additional logarithmic Hamiltonians  can be expressed by
   the logarithm of the Lax operator and are well-defined only after suitable constraint
 included in twistor data.

 Finally, from the genus zero free energy of the Benney hierarchy, we compute the associated
 $G$-function to construct the genus one free energy and then quantize the Poisson brackets
  of the Benney hierarchy  using the Dubrovin-Zhang's (DZ) approach to bi-Hamiltonian
  structure in 2d TFT \cite{DZ}. On the other hand, we also``quantize" these
   Poisson structures from the Oevel-Strampp's (OS) brackets of the Kaup-Broer (KB)
    hierarchy \cite{KO,OS}.
 We find that after appropriate differential substitutions, they are matched up to
  genus one correction.

   The paper is organized as follows. In next section we compute the primary free
    energy from the bi-Hamiltonian structure of the Benney hierarchy and then introduce
     the additional logarithmic flows commuting with the ordinary Benney flows.
      In Sec. III we compute topological correlation functions using the LG formulation
    and show that the Benney hierarchy can be derived from genus zero topological
     recursion relations. Sec. IV is devoted to finding the twistor data to
    establish the string equation including the additional logarithmic flows.
    In Sec. V we show that the genus one correction of the Poisson brackets
    obtained by DZ quantization coincide with the OS brackets
     after appropriate differentiable substitutions of the dynamical variables.
    In the last section we discuss some problems to be investigated.

\section{Bi-Hamiltonian structure and free energy}
In this section, we shall investigate the relations between the bi-Hamiltonian structure and
 its associated free energy of the Benney hierarchy.
  It turns out that we can find additional logrithmic hierarchy using the free energy,
  which together with the ordinary hierarchy (\ref{ben1}) form the flows of
gravitational  couplings in TFT of genus zero, the topic to be discussed in next section.

  The bi-hamiltonian structure of the Benney hierarchy (\ref{ben1}) is given by
 \cite{Li,CT2}
\[
 \frac{\pa {\mathbf{v}}}{\pa T_n}=J_1\frac{\de H_{n+1}}{\de {\mathbf{v}}}=
 J_2\frac{\de H_n}{\de {\mathbf{v}}},
\]
\be
J_1=
 \left(
 \ba{cc} 0 & \pa \\
  \pa & 0
   \ea
    \right),\qquad
 J_2=
 \left(
  \ba{cc} 2\pa & \pa v^1 \\
   v^1\pa & v^2\pa+\pa v^2
    \ea
 \right).
 \label{pb}
 \ee
with Hamiltonians defined by
\[
H_n=\frac{1}{n}\int \res  L^n,
\]
where $\pa\equiv \pa/\pa X$ and $\res L^n$ being the coefficient of $p^{-1}$ of $ L^n$.
 We list some of them as follows:
 \bea
  H_1&=&\int v^2,\no\\
  H_2&=&\int
 v^1v^2, \no\\
 H_3&=&\int \left[(v^1)^2v^2+(v^2)^2\right], \no\\
 H_4&=&\int \left[(v^1)^3v^2+3v^1(v^2)^2\right]. \no
  \eea
{\bf Remarks 2.} The second bracket $J_2$ in fact reveals the classical limit of Virasoro-$U(1)$-Kac-Moody
 algebra \cite{CT2} with $v^2$ being the Diff$S^1$ tensor of weight 2 and $v^1$ a tensor
  of weight 1. This is due to the fact that the Diff$S^1$ flows are just the Hamiltonian flows
   generated by the Hamiltonian $H_1=\int v^2$.

 Besides the concept of integrability, the geometrical means of the Poisson
 brackets (\ref{pb})  is profound.
 The essential idea is based on the fact that the bi-Hamiltonian structure $J_1$ and $J_2$
  can be written as
   \bean
    J_1 &=&
 \left(
 \ba{cc} 0 & 1 \\
  1 & 0
   \ea
    \right) \pa \equiv \eta^{\alpha\beta} \pa,  \no \\
 J_2 &=&
 \left(
  \ba{cc} 2 &  v^1 \\
   v^1 & 2v^2
    \ea
 \right) \pa +\left(
  \ba{cc} 0 &  1 \\
   0& 0
    \ea
 \right)v^1_{X}+ \left(
  \ba{cc}0&  0 \\
   0 & 1
    \ea
 \right)v^2_{X} \equiv g^{\alpha\beta}(v)\pa +\Ga_{\gamma}^{\alpha\beta}(v)v^\gamma_{ X}. \no
\eean
 where $\Gamma^{\alpha\beta}_\gamma(v)$ is the contravariant Levi-Civit\`a connection
  of the contravariant flat metric $g^{\alpha\beta}(v)$. Both $J_1$ and $J_2$ are Poisson
   brackets of hydrodynamic type introduced by Dubrovin and Novikov \cite{DN1}.
    The existence of a bi-Hamiltonian structure means that $J_1$ and $J_2$ have
   to be compatible, i.e, $J=J_1+\lambda J_2$ must be a Hamiltonian structure as well for
all value of $\lambda$. The geometric setting of this bi-Hamiltonian
  structure of hydrodynamic system is provided by  Frobenius manifolds\cite{Du2,Du3}.
   One way to define such manifolds is to construct a function
   $F(t^1, t^2, \cdots, t^m)$ such that the associated functions,
\be
c_{\alpha\beta\gamma}= \frac{\pa^3 F(t)}{ \pa t^\alpha \pa t^\beta \pa t^\gamma},
 \label{df}
 \ee
 satisfy the following conditions \cite{Du2}:
\begin{itemize}
\item
 The matrix $\eta_{\alpha\beta}=c_{1\alpha\beta}$ is constant and non-degenerate.
 (for the discussion of degenerate cases, see \cite{S})
\item The functions $c_{\beta\gamma}^{\alpha}=\eta^{\alpha\epsilon}c_{\epsilon\beta\gamma}$
 define an associative commutative algebra with a unity element. The associativity will give
 a system of non-linear PDE for $F(t)$
 \be
\frac{\pa^3 F(t)}{\pa t^{\alpha} \pa t^{\beta} \pa t^{\lambda}} \eta^{\lambda \mu}
 \frac{\pa^3 F(t)}{\pa t^{\mu} \pa t^{\ga}
\pa t^{\sigma}} = \frac{\pa^3 F(t)}{\pa t^{\alpha} \pa t^{\ga} \pa t^{\lambda}}
 \eta^{\lambda \mu} \frac{\pa^3 F(t)}{\pa
t^{\mu} \pa t^{\beta} \pa t^{\sigma}}.
 \label{WDVV}
 \ee
\item The functions $F$ satisfies a quasi-homogeneity condition, which may be expressed
as
\[
{\mathcal L\/}_{E} F= d_{F}F+ (\mbox{quadratic~ terms}),
  \]
where $E$ is known as  the Euler vector field.
\end{itemize}
Equations (\ref{WDVV}) constitute the WDVV equations \cite{W2,DVV} arising from
 TFT (see Sec. III).
 A solution of the WDVV equations will be called primary free energy. Given any solution
 of the WDVV equation, one can construct a Frobenius manifold ${\mathcal M\/}$
  associated with it. On such a manifold one may interpret $\eta^{\alpha\beta}$ as
   a flat metric and $t^{\alpha}$ the flat coordinates. The associativity can be used
   to defines a Frobenius algebra on each tangent space $T^t {\mathcal M\/}$. This
  multiplication will be denoted by $u \cdot v$.
   Then one may introduce a second flat metric on ${\mathcal M\/}$ defined by
 \bea
  g^{\alpha\beta}=E(dt^\alpha\cdot  dt^\beta), \label{sec}
   \eea
 where $dt^\alpha\cdot  dt^\beta=c_{\gamma}^{\alpha\beta}dt^{\gamma}=
 \eta^{\alpha \sigma}c_{\sigma\gamma }^{\beta}
 dt^{\gamma}$. This metric, together
with the original metric $\eta^{\alpha\beta}$, define a flat pencil
 (i.e, $\eta^{\alpha\beta}+\lambda g^{\alpha\beta}$ is flat for any value of
$\lambda$). Thus, one automatically obtains a bi-Hamiltonian
 structure from a Frobenius  manifold ${\mathcal M\/}$.
 The corresponding Hamiltonian densities are defined recursively by the formula \cite{Du2}
 \bea
  \frac{\pa^2 h_{\alpha}^{(n)}}{\pa t^\beta \pa
t^\gamma}=c_{\beta\gamma}^\sigma \frac{\pa h_{\alpha}^{(n-1)}}{\pa t^\sigma}, \label{rec}
 \eea
  where $n \geq 1, \alpha =1,2, \cdots, m,$ and $h_{\alpha}^{(0)}=\eta_{\alpha \beta}t^{\beta}$.
   The integrability conditions for this systems are automatically satisfied when
the $c_{\alpha\beta}^\gamma$ are defined as above.

 \indent To find the free energy associated with the Benney hierarchy (\ref{ben1}),
  one might set $t^1=v^1=h_2^{(0)}, t^2=v^2=h_{1}^{(0)}$ and thus
 \bea
  \eta^{\alpha\beta}(t)=
 \left(
 \ba{cc} 0 & 1 \\
  1 & 0
   \ea
    \right),\qquad
 g^{\alpha\beta}(t)=
 \left(
  \ba{cc} 2 & t^1 \\
   t^1& 2t^2
    \ea
 \right). \label{met}
\eea
 We remark that the flat metrics $\eta^{\alpha\beta}(t)$ and $g^{\alpha\beta}(t)$
  satisfy $\eta^{\alpha\beta}(t)=\pa g^{\alpha\beta}(t)/\pa t^1$ and $\int t^1$ and $\int t^2$
   turn out to be the Casimirs for the bi-Hamiltonian structure of the hierarchy.
 In fact, those $c_{\alpha\beta}^\gamma$ can be determined by (\ref{rec}) and (\ref{met}).
  For $\alpha=1$
we have
\[
 \int h_1^{(n)}=\frac{H_n}{(n-1)!}=\frac{1}{n!}\int\res L^n, \qquad
(h_1^{(0)}=t^2)
 \]
which, up to a normalization, are the Hamiltonian densities of the Benney hierarchy and
  \bea
    c^1_{11}=1,\quad c^1_{12}=c^1_{21}=0,\quad c^1_{22}=\frac{1}{t^2}, \no \\
c^2_{11}=c^2_{22}=0, \quad c^2_{21}=c^2_{12}=1. \label{con}
 \eea

  Then, from (\ref{con}) and (\ref{df}), we get  immediately the free energy
\bea
 F(t^1,t^2)=\frac{1}{2}(t^1)^2t^2+\frac{1}{2}(t^2)^2\left(\log t^2-\frac{3}{2}\right).
  \label{free}
 \eea
  Also, from (\ref{sec}), (\ref{met}) and (\ref{con}), we can obtain the
  associated Euler vector field
\[
E=t^1\frac{\pa }{\pa t^1}+2t^2\frac{\pa}{\pa t^2}
 \]
  which implies the quasi-homogeneity condition:
\[
{\mathcal{L}}_E F(t)=4 F(t)+(t^2)^2.
 \]

   Next, we turn to the hierarchy corresponding to $\alpha=2$ with
$h_2^{(0)}=t^1=v^1$. Using (\ref{rec}) and (\ref{con}), we get
 \bea
 {h}_{2}^{(1)} &=& \frac{(t^1)^2}{2}+t^2(\log t^2-1),\no\\
 {h}_{2}^{(2)} &=& \frac{(t^1)^3}{6}+t^1t^2(\log t^2-1),\no\\
 {h}_{2}^{(3)} &=& \frac{(t^1)^4}{24}+\frac{1}{2}(t^1)^2t^2(\log t^2-1)+
 \frac{1}{2}(t^2)^2\left(\log t^2-\frac{5}{2}\right),\no\\
 {h}_2^{(4)} &=& \frac{(t^1)^5}{120}+\frac{1}{6}(t^1)^3t^2(\log
 t^2-1)+\frac{1}{2}t^1(t^2)^2\left(\log t^2-\frac{5}{2}\right). \no
 \eea
Motivated by the work of \cite{EY,EHY}, the Hamiltonian densities
 $h_2^{(n)}$ can be expressed as
\[
{h}_{2}^{(n)} =\frac{2}{n!}\res [L^n(\log L-c_n)],
 \]
  with the prescription
   \bea
  \log  L &=&\log(p+t^1+t^2p^{-1}) \no\\
  &=&\frac{1}{2}\log t^2+\frac{1}{2}\log (1+t^1p^{-1}+t^2p^{-2})+
\frac{1}{2}\log \left(1+\frac{t^1}{t^2}p+\frac{1}{t^2}p^{2}\right)
 \label{logl}
  \eea
  and $c_n=\sum_{j=1}^n\frac{1}{j},c_0=0$.
Then the Lax flows corresponding to $h_{2}^{(n)}$  are
\be
\frac{\pa  L}{\pa \bar{T}_n}=2 \{\bar{B}_n, L \},
 \qquad \bar{B}_n=[L^n(\log L-c_n)]_{\geq 1}
 \label{logflow}
\ee
 or, in terms of bi-Hamiltonian structure
\[
 \frac{\pa {\mathbf{v}}}{\pa \bar{T}_n}=J_1\frac{\de \bar{H}_{n+1}}{\de
{\mathbf{v}}}=
 J_2\frac{\de \bar{H}_n}{\de {\mathbf{v}}}, \]
where the Hamiltonians $\bar{H}_n$ are defined by
\[
\bar{H}_n=\frac{2}{n}\int \res [ L^n(\log L-c_n)]=(n-1)!\int h_2^{(n)}.
\]
These Hamiltonians generate additional flows which are compatible with the ordinary
 Benney flows. We will see later that it is these logrithmic flows (\ref{logflow}) which
  together with  the ordinary flow (\ref{ben1}) implies that the Benney hierarchy
   can be formulated as a  2d TFT coupled to gravity.

 \section{topological string at genus zero}

 In this section we would like to set up the correspondence between the Benney hierarchy
  and its associated TFT at genus zero. Let us first recall some
   basic notions in TFT.

 A topological matter theory can be characterized by a set of BRST invariant observables
 $\{\op_1, \op_2,...\}$ with couplings $\{T^\alpha\}$ where $\op_1$ denotes the
  identity operator. If the number of observables is finite the theory is called
  topological  minimal model and the observables are refereed to the primary fields.
  When the theory couples to gravity, a set of new observables emerge as gravitational
 descendants $\{\sigma_n(\op_\alpha), n=1,2,\cdots\}$ with new coupling constants
  $\{T^{\alpha,n}\}$.  The identity operator $\op_1$ now becomes the puncture operator $P$.
   For convenience we can identify the primary fields  $\op_\alpha$ and the
   coupling constants $T^\alpha$ to  $\sigma_0(\op_\alpha)$ and $T^{\alpha,0}$, respectively.
 As usual, we shall call the space spanned by $\{T^{\alpha,n}, n=0,1,2,\cdots\}$ the full
 phase space  and  the subspace parametrized by $\{T^\alpha\}$ the small phase space.
 These coupling times decribe the perturbative flows with respect to the corresponding
 critical theory (in which $T^{\alpha,n}=0$).

 For a topological model the most important quantities are correlation functions which
 describe the topological properties of the manifold where the model lives.
  The generating function of correlation functions is the full free energy defined by
 \be
{\mathcal F\/}(T)=\sum_{g=0}^\infty {\mathcal F\/}_g(T)=\sum_{g=0}^\infty
 \lan e^{\sum_{\alpha,n} T^{\alpha,n}\sigma_n(\op_{\alpha})}\ran_g
 \label{gfree}
 \ee
 where $\lan\cdots\ran_g$ denotes the expectation value on a Riemann surface of
  genus $g$ with respect  to a classical action.
   In the subsequent sections, we will omit the exponential factor
 without causing any confusion. Therefore a generic $m$-point correlation function
  can be calculated as follows
\be
 \lan\sigma_{n_1}(\op_{\alpha_1})\sigma_{n_2}(\op_{\alpha_2})\cdots\sigma_{n_m}
 (\op_{{\alpha}_m})\ran_g=\frac{\pa^m {\mathcal F\/}_g}{\pa T^{\alpha_1,n_1}
 \pa T^{\alpha_2,n_2}\cdots\pa T^{\alpha_m,n_m}}.
 \label{defcorr}
 \ee
 In the following, we shall restrict ourselves to the trivial topology, i.e.,
  the genus zero sector ($g=0$) since this part is more relevant to dispersionless
   integrable hierarchies. In particular, the genus zero free energy restricting
   on the small phase space is the primary free energy defined by
   \be
 {\mathcal F\/}_0|_{T^\alpha=t^\alpha, T^{\alpha,n\geq 1}=0}=F(t).
   \ee

 Let us define some genus zero correlation functions on the small phase space. The metric on the
 space of primary fields is defined by
 \be
\lan P\op_{\alpha}\op_{\beta}\ran=\eta_{\alpha\beta}. \label{metric}
 \ee
When $\eta_{\alpha\beta}$ is independent of the couplings we call it the flat metric and
  the couplings $T^\alpha$ the flat coordinates. In fact, a three-point function
   in the small phase space can be expressed as
\be
\lan \op_{\alpha}\op_{\beta}\op{\gamma}\ran=\frac{\pa F}
 {\pa T^{\alpha}\pa T^{\beta}\pa T^{\gamma}} \equiv c_{\alpha\beta\gamma}
 \label{ctopo}
 \ee
which provide the structure constants of the commutative associative algebra
\be
\op_{\alpha}\op_{\beta}=c_{\alpha\beta}^\gamma \op_\gamma
 \ee
with constraints $ c_{\alpha\beta}^\gamma=\eta^{\gamma\sigma}c_{\sigma\alpha\beta},
 c_{1\alpha\beta}=\eta_{\alpha\beta} $.
 The associativity of $c_{\alpha \beta}^{\ga}$, i.e.,
\[
c_{\alpha \beta}^{\mu}c_{\mu \ga}^{\sigma}=c_{\alpha \ga}^{\mu}c_{\mu \beta}^{\sigma},
\]
 will give the WDVV equations (\ref{WDVV}).

 Now, Let's return to the Benney hierarchy. Since the Benney hierarchy is a two-variable theory,
  thus only two primary fields $\{\op_1=\op_P\equiv P, \op_2=\op_Q\equiv Q\}$ are involved in the
  TFT formulation and  we shall identify $v^1|_{T^{\alpha,n\geq 1}=0}=T^P$ and
   $v^2|_{T^{\alpha,n\geq 1}=0}=T^Q$ on the small phase space.
   Therefore the Lax operator   in small phase space is written as $L(z)=z+T^P+T^Qz^{-1}$
    which can be viewed as a superpotential in LG formulation of TFT
     \cite{V,DVV}.
     According to the LG theory, the primary fields are defined by
  \be
 \op_P(z)=\frac{\pa L(z)}{\pa T^P}=1,\qquad
 \op_Q(z)=\frac{\pa L(z)}{\pa T^Q}=z^{-1}
  \ee
which can be used to compute the three-point correlation functions through the formula
 \cite{V,DVV}:
\be
c_{\alpha\beta\gamma}=\res_{L'=0}\left[
 \frac{\op_\alpha(z)\op_\beta(z)\op_\gamma(z)}{\pa_zL(z)}\right].
 \label{3pt}
\ee
 It is easy to show that (\ref{3pt}) reproduces the previous $c_{\alpha\beta\gamma}$ and $F$ on
  the small phase space. In particular, the flat metric on the space of primary fields is given by
\[
\eta_{PQ}=\eta_{QP}=1,\qquad \eta_{PP}=\eta_{QQ}=0
\]
as we obtained previously. Now we can impose the fundamental correlation functions as
 $\langle PP\rangle=\pa^2F/\pa(T^P)^2=T^Q$,
  $\langle PQ\rangle=\pa^2F/\pa T^P\pa T^Q=T^P$, and
    $\langle QQ\rangle =\log T^Q$. Although these two-point correlation functions are defined
    on the small phase space, however, it has been shown \cite{DW} that they can be defined
     on the full phase space through the variables $v^1$ and $v^2$ in which the
      gravitational couplings $T^{\alpha,n}$ do not vanish.
    Hence it is easy to write down these genus zero two-point functions on the full phase
     space and obtain the following constitutive relations:
 \bea
\langle PP\rangle&=&\frac{\pa^2{\mathcal F\/}_0}{\pa (T^P)^2}=v^2,\no\\
 \langle PQ\rangle&=&\frac{\pa^2{\mathcal F\/}_0}{\pa T^P\pa T^Q}=v^1,
 \no\\
\langle QQ\rangle &=&\log\lan PP\ran \label{consti}
 \eea
which will be important to provide the connection between the Benney hierarchy and its associated
 TFT.\\
 {\bf Remark 3.\/} In general, for the two-primary models,
  $\langle QQ\rangle=f(\langle PP\rangle)$ where the function $f(x)$ is model dependent \cite{DW}.
  In \cite{BX}, the same relation $f(x)=\log x$ has been imposed to extract the nonlinear
  Schr\"odinger hierarchy from the hermitian one-matrix model at finite $N$. However, for
  the $CP^1$ model \cite{W2,DW}, $f(x)=e^x$.

  Based on the constitutive relations (\ref{consti}),
  we can identify the gravitational flows for $v^1$ and $v^2$ in the full phase space
  as the Lax flows by taking into account the genus zero topological recursion
   relation \cite{W2,DW}:
\be
\langle \sigma_n(\op_\alpha)AB\rangle=\sum_{\beta,\gamma=P,Q}n\langle \sigma_{n-1}(\op_\alpha)
 \op_\beta\rangle \eta^{\beta\gamma}\langle \op_\gamma AB\rangle, \qquad \alpha=P,Q.
 \label{recursion}
\ee
 For example, setting $n=1,~\op_\alpha=P$ and $A=P, B=Q$ then
 \bean
\frac{\pa v^1}{\pa T^{P,1}} &=&\langle \sigma_1(P)PQ\rangle\\
 &=&\langle PP\rangle\langle QPQ\rangle+\langle PQ\rangle\langle PQQ\rangle\\
 &=& \langle PP\rangle\lan QQ\ran'+\lan PQ\ran\lan PQ\ran'\\
 &=&\left[\frac{1}{2}(v^1)^2+v^2\right]'
 \eean
where we denote $f'=\pa f/\pa T^P=\pa f/\pa X$ . Similarly, taking $A=P,~B=P$ we have
\[
\frac{\pa v^2}{\pa T^{P,1}}=\lan \sigma_1(P)PP\ran=(v^1v^2)'.
\]
On the other hand, taking $n=2$ we get
 \bea
 \frac{\pa v^1}{\pa T^{P,2}} &=& \left[\frac{1}{3}(v^1)^3+2v^1v^2\right]',\no\\
 \frac{\pa v^2}{\pa T^{P,2}} &=& \left[(v^1)^2v^2+(v^2)^2\right]'.
 \label{pflow}
 \eea
Comparing the above equations with the Lax flows (\ref{ben2}), we shall identify
 $T_n=T^{P,n-1}/n,~(n=1,2,\cdots)$.

Next, let us turn to the $T^{Q,n}$ flows. Choosing $\op_\alpha=Q$ and using the topological
 recursion relation (\ref{recursion}), we obtain
 \bea
 \frac{\pa v^1}{\pa T^{Q,1}}&=&(v^1\log v^2)',\no\\
 \frac{\pa v^2}{\pa T^{Q,1}}&=&\left[\frac{1}{2}(v^1)^2+v^2(\log v^2-1)\right]',\no\\
  \frac{\pa v^1}{\pa T^{Q,2}}&=&\left[(v^1)^2\log v^2+2v^2(\log v^2-2)\right]',\no\\
   \frac{\pa v^2}{\pa T^{Q,2}}&=&\left[\frac{1}{3}(v^1)^3+2v^1v^2(\log v^2-1)\right]'
   \label{qflow}
 \eea
which are nothing but the Lax flows associated with the logarithmic
 operator $\bar{B}_n$ for $n=1,2$ under the identification
$\bar{T}_n=T^{Q,n}$.

It turns out that equations (\ref{pflow}) and (\ref{qflow}) can be recasted into the Lax form
  in terms of coupling times $T^{\alpha,n}$ as
 \bea
  \frac{\pa L}{\pa
T^{P,n}}&=&\frac{1}{n+1}\{L^{n+1}_{\ge 1}, L\},\no\\
 \frac{\pa L}{\pa T^{Q,n}}&=&2\{[L^n(\log L-c_n)]_{\ge 1}, L\}.
 \label{pqflow}
 \eea
 and the associated commuting Hamiltonian flows with respect to the bi-Hamiltonian structures
  are thus given by
 \bea
\frac{\pa {\mathbf{v}}}{\pa T^{P,n}}&=&\{H_{P,n+1}, {\mathbf{v}}\}_1=\{H_{P,n},{\mathbf{v}}\}_2,
 \qquad H_{P,n}=\frac{1}{n(n+1)}\int \res L^{n+1}\no\\
 \frac{\pa {\mathbf{v}}}{\pa T^{Q,n}}&=&\{H_{Q,n+1},
{\mathbf{v}}\}_1=\{H_{Q,n}, {\mathbf{v}}\}_2,\qquad H_{Q,n}=\frac{2}{n}\int \res [L^n(\log L-c_n)]
 \label{hflow}
 \eea
 where $n=0,1,2,\cdots$.\\
  {\bf Remark 4.\/} We notice that  (\ref{free}) corresponds to the primary free energy
 of the dKP system in the $W_{0,1}$-model \cite{AK} defined by the Lax operator $L=p+v_{-1}/(p-s)$
 under the identification $s=T^P, v_{-1}=T^Q$. They also show that another choice leads to
 the dToda system in the  $CP^1$ topological sigma model \cite{EY,EHY}.

Furthermore, using the constitutive relations (\ref{consti}) and the Lax flows (\ref{pqflow}), we
 have
 \bean
\frac{\pa v^1}{\pa T^{\alpha,n}}&=&\frac{\pa \lan\sigma_n(\op_\alpha)Q
 \ran}{\pa T^{P}}=(R^{(1)}_{\alpha,n})',\\
 \frac{\pa v^2}{\pa T^{\alpha,n}}&=&\frac{\pa \lan\sigma_n(\op_\alpha)P
 \ran}{\pa T^{P}}=(R^{(2)}_{\alpha,n})',
 \eean
where $R^{(\beta)}_{\alpha,n}$ are the analogues of the Gelfand-Dickey potentials \cite{Di} of the
 KP hierarchy, given by
 \bea
  R^{(1)}_{P,n}&=&\lan\sigma_n(P)Q
\ran=\frac{1}{n+1}(L^{n+1})_0,\no\\
 R^{(1)}_{Q,n}&=&\lan\sigma_n(Q)Q
\ran=2[L^n(\log L-c_n)]_0,\no\\
  R^{(2)}_{P,n}&=&\lan\sigma_n(P)P
\ran=\frac{1}{n+1}\res L^{n+1},\no\\
 R^{(2)}_{Q,n}&=&\lan\sigma_n(Q)P
\ran=2\res[L^n(\log L-c_n)]. \label{gdpot}
 \eea
 For example, the two-point correlators involving the first and the second descendants are
 \bean
    \lan\sigma_1(P)P\ran&=&v^1v^2,\\
    \lan\sigma_1(P)Q\ran&=&\frac{1}{2}(v^1)^2+v^2, \\
\lan\sigma_1(Q)P\ran&=&\frac{1}{2}(v^1)^2+v^2(\log v^2-1),\\
  \lan\sigma_1(Q)Q\ran&=&v^1\log v^2, \\
  \lan\sigma_2(P)P\ran&=&(v^1)^2v^2+(v^2)^2,\\
\lan\sigma_2(P)Q\ran&=&\frac{1}{3}(v^1)^3+
  2v^1v^2 ,\\
\lan\sigma_2(Q)P\ran&=&\frac{1}{3}(v^1)^3+2v^1v^2(\log v^2-1),\\
  \lan\sigma_2(Q)Q\ran&=&(v^1)^2\log v^2+2v^2(\log v^2-2).
\eean

 Finally, we would like to remark that the last two equations of (\ref{gdpot}) can
  be integrated to yield
 \bean
 \lan\sigma_n(P)\ran&=&\frac{1}{n+1}R^{(2)}_{P,n+1}=\frac{1}{(n+1)(n+2)}\res L^{n+2},\\
 \lan\sigma_n(Q)\ran&=&\frac{1}{n+1}R^{(2)}_{Q,n+1}=\frac{2}{n+1}\res[L^{n+1}(\log L-c_{n+1})].
 \eean
which are just the LG representation for one-point functions of gravitational decendants
 at genus zero.


\section{The twistor data and string equation}
In this section we would like to discuss the string equation of the 2d TFT associated with
 the Benney hierarchy, which govern the dynamics of the variables $v^\alpha$ (or fundamental
 correlators) in the full phase space.
 The Lax formulation in Sec. II for Benney hierarchy is in fact similar to the formulation of the
 dToda type hierarchy  \cite{TT,TT1}. Based on this observation,
 we can reproduce the Benney equations by imposing constraints on the Lax operators
 and the associated Orlov operators of the dmKP hierarchy \cite{CT1}
 through the twistor data (see below).
 We shall remark, however, that the flow equations corresponding to $h_{2}^{(0)}$
 are absent in the standard  formulation of the dToda type hierarchy.
  So we have to properly extend the standard Orlov operator
 to include the additional hierarchy equations (\ref{logflow}).
 We will follow closely that of \cite{Kh} to show that the constraints imposing on
 the twistor data  implies the string equation of the Benney hierarchy.

\indent Let us consider two Lax operators $ \mu$ and $\tilde{ \mu}$ with the following
 Laurent expansions
 ($T^1=X$):
  \bea
   \mu &=& p+\sum_{n=0}^{\infty} v^{n}( T,  \tilde{ T}) p^{-n}, \no \\
    {\tilde{ \mu}}^{-1} &=& \tilde {v}_{0}(T, \tilde {T})
p^{-1}+\sum_{n=0}^{\infty} \tilde {v}_{n+1}(T, \tilde {T})p^n . \label{pole}
 \eea
  which satisfy the commuting Lax flows
   \bea
    \frac{\pa \mu}{\pa T_n} &=& \{B_n, \mu \}, \qquad  \frac{\pa \mu}{\pa \tilde {T}_n}
     = \{\tilde {B}_n, \mu \}, \no \\
      \frac{\pa \tilde {\mu}}{\pa T_n} &=& \{B_n, \tilde {\mu} \}, \qquad
       \frac{\pa \tilde {\mu}}{\pa \tilde {T}_n} = \{\tilde {B}_n, \tilde {\mu} \},\quad (n=1,2,3
\cdots) \label{bar1}
 \eea
  where the Poisson bracket $\{,\}$ is defined as before and
\[
B_n\equiv (\mu^n)_{\geq 1}, \qquad \tilde {B}_n\equiv ({\tilde {\mu}}^{-n})_{\leq 0}.
 \]
  Next, we consider the Orlov operators corresponding to dmKP
 \bea
  M &=& \sum_{n=1}^{\infty}nT_n
\mu^{n-1}+\sum_{n=1}^{\infty}w_n(T,\tilde {T})\mu^{-n} \no  \\
 \tilde {M} &=&- \sum_{n=1}^{\infty} n \tilde{T}_n {\tilde {\mu}}^{-n-1}+X+\sum_{n=1}^{\infty}
 \tilde {w}_n (T,\tilde {T}){\tilde {\mu}}^{n}. \label{or}
   \eea
  with constraint
    \bea
\{\mu, M\}=1, \qquad \{\tilde {\mu}, \tilde{ M} \}=1.  \label{unit}
 \eea
  In fact, the coefficient functions $w_n$ and $\tilde {w}_n$ in the Orlov operators
   are defined by the above canonical relations and the following flow equations
 \bea
  \frac{\pa M}{\pa T_n} &=& \{B_n, M \},
\qquad  \frac{\pa M}{\pa \tilde{T}_n} = \{\tilde{B}_n, M \}, \label{mflow} \\
 \frac{\pa \tilde{M}}{\pa T_n} &=& \{B_n, \tilde{M} \}, \qquad \frac{\pa \tilde {M}}
 {\pa \tilde{T}_n}
 = \{\tilde{B}_n, \tilde{M} \}.\quad (n=1,2,3 \cdots) \label{bar2}
   \eea
   Inspired by the twistor construction (or Riemann-Hilbert problem) for the solution
    structure of the dToda hierarchy \cite{TT,TT1}, we now give the twistor construction
     for the Benney hierarchy.
\begin{theorem}
 Let  $f(p,X),g(p,X), \tilde {f}(p,X), \tilde
{g}(p,X)$ be functions satisfying
 \bea
  \{f(p,X), g(p,X)\}=1, \qquad \{\tilde {f}(p,X), \tilde {g}(p,X)\}=1.
 \label{pois}
  \eea
   Then the functional equations
    \bea
     f(\mu, M)=\tilde {f}(\tilde {\mu}, \tilde {M}),
    \qquad g(\mu, M)=\tilde {g} (\tilde {\mu}, \tilde {M})
     \label{fun}
\eea
 can get a solution of (\ref{bar1}) and (\ref{bar2}).
  We call the pairs $(f,g)$ and $(\tilde {f}, \tilde {g})$ the twistor data of the solution.
\end{theorem}
The proof is provided in appendix A.

  To reduce the above Theorem to the Benney hierarchy, we have
 to impose  the following constraint on the Lax operators
 \bea
  L=\mu={\tilde {\mu}}^{-1}, \label{tw1}
   \eea
 that is, $f(p,X)=p$ and $\tilde {f}(p,X)=p^{-1}$.
  As a result, the time variables $\tilde {T}_n$
  can be eliminated via the following identification:
      \bean
       \tilde {T}_n=-T_n. \no
        \eean
From (\ref{pois}), the twistor data $g(p,X)$ and $\tilde {g}(p,X)$ can be
 assumed the following form
\be
g(p,X)=X-\sum_{n=2}^{\infty}nT_n p^{n-1}, \qquad \tilde {g}(p,X)=-Xp^2.
 \label{gfunc}
 \ee
where the second part of $g(p,X)$ is responsible for the string equations (see below).
 By the Theorem 1 and equation (\ref{gfunc}), we get the following constraint
  for the Orlov operators:
  \bea
   M- \sum_{n=2}^{\infty}nT_n {\mu}^{n-1}=-{\tilde {\mu}}^2 \tilde {M}.
\label{tw2}
 \eea
  It's the above constraint that leads to the string equations.

  So far, the twistor construction only involves the Lax flows (\ref{ben1}). To obtain the string
  equations associated with the TFT described in Sec. II, we
  have to modify the Orlov operator $M$ to include the additional flows (\ref{logflow}). Namely,
   it's necessary to introduce the flows generated by the logarithmic operator
\[
 \bar {B_n}=[L^n(\log L-c_n)]_{\geq 1}.
  \]
where we have imposed the constraint (\ref{tw1}) and used the prescription of the series
  expansion (\ref{logl}) for $\log L$.
   Let $ \bar {T_n} $ be the time variables of additional flows generated by $\bar {B_n}$
 then the Orlov operator $M$ is deformed  by these new flows to $M'$ so that
 \bea
  \frac{\pa M'}{\pa \bar {T_n}}=2 \{\bar {B_n} ,M' \}. \label{tw3}
   \eea
   To construct the modified Orlov operator $M'$, it is convenient to using the dressing
    method \cite{TT1}. Let us first express the original Lax operator
$\mu$ and its
     conjugate Orlov operator $M$ in dressing form (similarity
transformation)
 \be
 \mu = e^{\mbox{ad}\Th}(p),\qquad M=e^{\mbox{ad}\Th}\left(\sum_{n=1}^{\infty} nT_n p^{n-1}\right)
    \label{dress1}
    \ee
    where
    \[
 e^{\mbox{ad}f}(g)=g+\{f,g\}+\frac{1}{2!} \{f, \{f,g\} \}+\cdots.
    \]
     One can understand that this is the canonical transformation generated by $\Th(T,\bar{T},p)$
     (for the $\bar{T}$-dependence, see below) and its flow equations (Sato equations)
      can be written as \cite{TT1}
  \bea
   \nabla_{T_n,\Th}\Th=B_n-e^{\mbox{ad}\Th}(p^n).
    \label{sato}
    \eea
    where
    \[
    \nabla_{T_n,\Th}\Th \equiv \sum_{k=0}^{\infty}\frac{1}{(k+1)!}(\mbox{ad}\Th)^k
    (\frac{\pa \Th}{\pa T_n}).
    \]
    It is easy to show that (\ref{sato}) together with (\ref{dress1})
     implies the flow equations (\ref{mflow}).

        In contrast with equation (\ref{sato}), the $\bar {T_n}$ flows for $\Th$ are given by
\[
   \nabla_{\bar{T}_n,\Th}\Th=2\bar{B}_n-e^{\mbox{ad}\Th}[p^n(\log p-c_n)].
   \]
and a similar argument reaches the modified Orlov operator
 \bea
  M'&=& e^{\mbox{ad}\Th} \left(\sum_{n=1}^{\infty}n T_n p^{n-1}+2
\sum_{n=1}^{\infty} n \bar {T_n}p^{n-1}(\log p-c_{n-1}) \right )\no\\
 &=& \sum_{n=1}^{\infty}n T_n
{\mu}^{n-1}+\sum_{n=1}^{\infty} w_n {\mu}^{-n}
   + 2 \sum_{n=1}^{\infty} n \bar {T}_n{\mu}^{n-1}(\log \mu -c_{n-1})
   \label{modm}
 \eea
 which  satisfies the additional flow equations (\ref{tw3}).

 Now we are in the position to derive the string equation.
Following \cite{Kh} we decompose $M'$ into the positive power part $(M')_{\geq 1}$ and the
 non-positive power part $(M')_{\leq 0}$ by using equations (\ref{tw2}) and (\ref{modm}).
  It turns out  that
 \bean
  \left(M'-\sum_{n=2}^{\infty}nT_n {\mu}^{n-1} \right)_{\geq 1}
   &=&  2 \left [\sum_{n=1}^{\infty} n \bar {T_n} {\mu}^{n-1}(\log \mu
-c_{n-1}) \right]_{\geq 1} \no \\
 &=& 2\sum_{n=1}^{\infty}n \bar{T_n} \bar B_{n-1}, \no \\
  \left(M'-\sum_{n=2}^{\infty}nT_n {\mu}^{n-1} \right)_{\leq 0}
  &=& -\left( {\tilde {\mu}}^2 \tilde {M} \right)_{\leq 0}\\
   &=& \left[ \sum_{n=1}^{\infty} n\tilde{T}_n {\tilde  {\mu}}^{-n+1}-X {\tilde {\mu}}^2
    -\sum_{n=1}^{\infty}\tilde {w}_n{\tilde {\mu}}^{n+2} \right]_{\leq 0}  \\
 &=& -\left( \sum_{n=1}^{\infty} n{T_n} {\mu}^{n-1} \right)_{\leq 0}
  \eean
 where, by the definition (\ref{pole}), we have used the fact that
  $(\mu^{-n})_{\geq 1}=(\tilde{\mu}^{n})_{\leq 0}=0$ for $n\geq 1$. Hence
    \bea
     g(\mu,M')= 2\sum_{n=1}^{\infty}n \bar{T_n} \bar B_{n-1}-\left( \sum_{n=1}^{\infty}
     n {T_n} {\mu}^{n-1} \right)_{\leq 0}
      \label{str}
      \eea
       which together with the canonical commutation relation $\{g(L, M'), L \}=-1$ implies
         \be
   \sum_{n=2}^{\infty} nT_n \frac{\pa L}{\pa T_{n-1}}+\sum_{n=1}^{\infty} n \bar{T_n}
    \frac{\pa  L}{\pa {\bar T}_{n-1}}=-1. \label{str1}
     \ee
     After shifting $ T_1\to  T_1-1$ and making the identification
      $T_n=T^{P,n-1}/n$ and ${\bar T}_n=T^{Q,n}$ as described before, we have
      \be
 \frac{\pa L}{\pa T^P}=1+\sum_{n=1}^{\infty} nT^{P,n}
  \frac{\pa L}{\pa T^{P,n-1}}+\sum_{n=1}^{\infty} n T^{Q,n}
    \frac{\pa  L}{\pa T^{Q,n-1}}.
    \label{prestr}
      \ee
Now taking the zero-th order term of  the  above equation and using the
 constitutive relation (\ref{consti}) yield
 \be
 \frac{\pa \lan PQ\ran}{\pa T^P}=1+\sum_{n=1}^{\infty}\sum_{\alpha=P,Q} nT^{\alpha,n}
  \lan\sigma_{n-1}({\mathcal O}_\alpha)PQ\ran.\no
      \ee
By integrating the above equation we obtain the universal string equation at genus zero \cite{DW}
 \bea
v^1(T)&=&T^P+\sum_{n=1}^{\infty}\sum_{\alpha=P,Q} nT^{\alpha,n}
  \lan\sigma_{n-1}({\mathcal O}_\alpha)Q\ran,\no\\
v^2(T)&=&T^Q+\sum_{n=1}^{\infty}\sum_{\alpha=P,Q} nT^{\alpha,n}
  \lan\sigma_{n-1}({\mathcal O}_\alpha)P\ran
\eea
 which describe the dynamics of $v^{\alpha}$ associated with the gravitational background.


\section{genus one correction of Poisson brackets}

In this section, we will compute the genus one correction to the
 bi-Hamiltonian structure  of the Benney hierarchy by using the DZ
  approach to integrable hierarchy associated with TFT \cite{DZ},
  which  consists of the following two main ingredients:
\begin{itemize}
\item introducing slow spatial and time variables scaling
\be
 T^{\alpha, n} \to \ep T^{\alpha,n},\qquad n=0,1,2,...
 \label{scaling}
 \ee
\item changing the full free energy as
\[ {\mathcal F\/} \to \sum_{g=0}^{\infty} \ep^{2g-2} {\mathcal F\/}_g,
 \]
\end{itemize}
where $\ep$ is the parameter of genus expansion.
 Thus all of the corrections  become series in $\ep$.

  To get a unambiguous genus one correction of the Hamiltonian flows (\ref{hflow}) one
  may expand the flat coordinates up to the $\ep^2$ order as
  \be
 t_\alpha=t_\alpha^{(0)}+\ep^2t_\alpha^{(1)}+O(\ep^4),\qquad t_\alpha=\eta_{\alpha\beta}t^\beta
   \ee
 where $t_\alpha^{(0)}$ is the ordinary Benney variables $v_\alpha$ satisfying
  (\ref{hflow}) and $t_\alpha^{(1)}$ is the genus one correction defined by
 \be
t_\alpha^{(1)}=\frac{\pa^2{\mathcal F\/}_1(T)}{\pa T^\alpha\pa X}.
 \ee
Then there exists a unique  hierarchy flows of the form \cite{DZ}
 \bea
  \frac{\pa t^\alpha}{\pa T^{\beta,n}}&=&\{t^\alpha(X), H_{\beta,n+1}\}_1+O(\ep^4),\no\\
  &=&\{t^\alpha(X), H_{\beta,n}\}_2+O(\ep^4)
  \label{1hflow}
 \eea
with
 \bea
 \{t^\alpha (X), t^\beta (Y)\}_i&=&\{t^\alpha (X), t^\beta (Y)\}^{(0)}_i+
 \ep^2\{t^\alpha (X), t^\beta (Y)\}^{(1)}_i+O(\ep^4),\qquad i=1,2\no\\
 H_{\alpha,n}&=&H_{\alpha,n}^{(0)}+\ep^2H_{\alpha,n}^{(1)}+O(\ep^4)
\eea
   That means under such correction the Poisson brackets $J_1$ and $J_2$ and the Hamiltonians
  will receive corrections up to $\ep^2$ such that the Hamiltonian flows
   (\ref{1hflow}) still commute with each other.

In \cite{DZ}, the genus-one part of the free energy has the form
 \bea
  {\mathcal F\/}_1(T)=\left[\frac{1}{24}
\log \det M^{\alpha}_\beta(t, \pa_X t)+ G(t)\right]_{t=v(T)}, \label{genus}
 \eea
  where the matrix $M^{\alpha}_\beta$ is given by
   \[
    M^{\alpha}_\beta(t, \pa_X t)= c^{\alpha}_{\beta\ga}(t) \pa_X t^{\ga},
 \]
  and $G(t)$ is a certain function satisfying the Getzler's equation \cite{G}.
   The first part of the formula (\ref{genus}) is quite simple on the small phase space,
    whereas, the second part describes, in the topological sigma-models,
 the genus one Gromov-Witten invariant of the target space and satisfies a complicated
  recursion relations \cite{G}. For the primary free energy (\ref{free}),
   the $G$-function satisfies the following simple ordinary differential eqaution \cite{DZ}:
\[
48 (t^2)^2 \frac{\pa^2 G}{\pa (t^2)^2}+ 24 t^2 \frac{\pa G}{\pa t^2}=2
 \]
 which can be easily solved as (up to a constant)
\be
G(t)=-\frac{1}{12} \log t^2. \label{gfunction}
 \ee
 The $G$-function of the Frobenius manifold also satisfies the quasi-homogeneity
  condition ${\mathcal L\/}_EG=-1/6$. The above $G$-function can also be derived using
  the tau function of the isomonodromy deformation problem arising in the
   theory of WDVV equations of associativity \cite{DZ}. We briefly describe the derivation
   in appendix B.

   On the other hand, a simple computation yields
 \bean
  M^{\alpha}_\beta=
 \left(
 \ba{cc} t_X^1 & t_X^2 \\
   \frac{t_X^2}{t^2} & t_X^1
   \ea
    \right) \no
\eean
 which together with (\ref{gfunction}) implies
  \[
   {\mathcal F\/}_1=\frac{1}{24} \log \left[(t_X^1)^2t^2-(t_X^2)^2\right]
   -\frac{1}{8}\log t^2.
 \]
  Using $c^\alpha_{\beta\gamma}$ and $ {\mathcal F\/}_1$ and consulting the procedure
   developed by Dubrovin and Zhang (c.f. Theorem 1, Theorem 2 and Proposition 3 in \cite{DZ}),
    after a straightforward but tedious computation, we obtain the genus one
     correction of the first Poisson bracket:
    \bea
  \{t^1(X),t^1(Y)\}_1&=&0 + O(\ep^4),  \no \\
  \{t^1(X),t^2(Y)\}_1&=& \de'(X-Y) + \frac{\ep^2}{6}
   \left[\frac{t^2_X}{(t^2)^2}\de''(X-Y)-\frac{1}{t^2}\de'''(X-Y)\right]+O(\ep^4), \no \\
  \{t^2(X),t^1(Y)\}_1&=& \de'(X-Y)
   + \frac{\ep^2}{6}\left[\left(\frac{t^2_{XX}}{(t^2)^2}-\frac{2(t^2_X)^2}{(t^2)^3}\right)
   \de'(X-Y)\right.\no\\
   &&\left.+\frac{2t^2_X}{(t^2)^2}\de''(X-Y)-\frac{1}{t^2}\de'''(X-Y)\right]+
   O(\ep^4), \no \\
 \{t^2(X),t^2(Y)\}_1&=& 0 +O(\ep^4). \label{cor1}
 \eea
On the other hand, for the second bracket, we get
 \bea
  \{t^1(X),t^1(Y)\}_2&=&2\de'(X-Y)
  +\frac{\ep^2}{12}\left[\left(\frac{t^2_{XX}}{(t^2)^2}-\frac{2(t^2_X)^2}{(t^2)^3}\right)
   \de'(X-Y)\right.\no\\
   &&+\left.\frac{3t^2_X}{(t^2)^2}\de''(X-Y)-\frac{2}{t^2}\de'''(X-Y) \right]+O(\ep^4),\no  \\
  \{t^1(X),t^2(Y)\}_2&=&  t^1_X\de(X-Y)+t^1\de'(X-Y)\no\\
  &&+\frac{\ep^2}{6}\left[\left(\frac{t^1t^2_X}{(t^2)^2}-\frac{t^1_X}{t^2}\right)
   \de''(X-Y)-\frac{t^1}{t^2}\de'''(X-Y)\right]+ O(\ep^4), \no \\
  \{t^2(X),t^1(Y)\}_2&=& t^1\de'(X-Y)
  +\frac{\ep^2}{6} \left[\left(\frac{t^1t^2_{XX}}{(t^2)^2}+\frac{2t^1_Xt^2_X}{(t^2)^2}
  -\frac{2t^1(t^2_X)^2}{(t^2)^3}-\frac{t^1_{XX}}{t^2}\right)
   \de'(X-Y)\right.\no\\
   &&+\left.\left(\frac{2t^1t^2_X}{(t^2)^2}-\frac{2t^1_X}{t^2}\right)\de''(X-Y)
    -\frac{t^1}{t^2}\de'''(X-Y)\right]+ O(\ep^4), \no \\
 \{t^2(X),t^2(Y)\}_2&=& t^2_X\de(X-Y)+2t^2\de'(X-Y)+ O(\ep^4). \label{cor2}
 \eea
Also, we can derive the genus-one corrections of the Hamiltonians, $H^{(1)}_{\alpha,n}$
 and some of them are:
  \bean
   H_{P,1}^{(1)} &=& -\int \left(\frac{t_X^1 t_X^2}{6 t^2} \right) dX,  \\
   H_{P,2}^{(1)} &=& -\int \left [\frac{(t_X^1)^2}{6 }
+\frac{t^1t_X^1 t_X^2}{6t^2} +\frac{(t_X^2)^2}{8t^2}  \right]  dX,  \\
 H_{Q,1}^{(1)} &=& -\int \left [\frac{(t_X^1)^2}{12t^2}
 +\frac{ (t_X^2)^2}{24(t^2)^2} \right]  dX,  \\
 H_{Q,2}^{(1)} &=& -\int \left [\frac{t^1(t_X^1)^2}{12t^2} +\left(\frac{1}{6t^2}+
 \frac{\log t^2}{6t^2}\right)t_X^1 t_X^2 +\frac{t^1(t_X^2)^2 }{24(t^2)^2} \right]  dX.
\eean
 Therefore the corresponding commuting flow equations up to genus-one corrections  are given by
  \bean
   \left(
 \ba{c} t^1 \\ t^2
   \ea
    \right)_{T_{P,0}} & = & \left(
 \ba{c} t^1 \\ t^2
   \ea
    \right)_X+O(\ep^4),  \\
\left(
 \ba{c} t^1 \\ t^2
   \ea
    \right)_{T_{P,1}} &=&
\left( \ba{c} \frac{(t^1)^2}{2}+t^2 \\ t^1 t^2  \ea \right)_X +\frac{\ep^2}{24}
 \left( \ba{c} \frac{2t_{XX}^2}{t^2}-\frac{3(t_X^2)^2}{(t^2)^2} \\
 4t_{XX}^1-\frac{4t_X^1 t_X^2}{t^2}  \ea \right)_X+O(\ep^4),  \\
\left(
 \ba{c} t^1 \\ t^2
   \ea
    \right)_{T_{Q,0}} &=& \left( \ba{c}  \log t^2 \\ t^1
\ea \right)_X + \frac{\ep^2}{24} \left( \ba{c} -\frac{2t_{XX}^2}{(t^2)^2}+
 \frac{2(t_{X}^1)^2}{(t^2)^2}+\frac{2(t_{X}^2)^2}{(t^2)^3} \\ 0 \ea \right)_X+O(\ep^4),  \\
  \left(
 \ba{c} t^1 \\ t^2
   \ea
    \right)_{T_{Q,1}} &=& \left( \ba{c} t^1 \log t^2 \\ \frac{1}{2}
(t^1)^2 +t^2( \log t^2 -1) \ea \right)_X  \no \\
 &&+ \frac{\ep^2}{24} \left( \ba{c} \frac{4t_{XX}^1}{t^2}
-\frac{2t^1t_{XX}^2}{(t^2)^2}+ \frac{2t^1(t_{X}^1)^2}{(t^2)^2}+
 \frac{2t^1(t_{X}^2)^2}{(t^2)^3}-\frac{6t_X^1 t_X^2}{(t^2)^2}  \\
  \frac{4t_{XX}^2}{t^2}- \frac{2(t_{X}^1)^2}{t^2}- \frac{5(t_{X}^2)^2}{(t^2)^2} \ea \right)_X
  +O(\ep^4).
   \eean

 Finally we would like to show that the genus one correction of the Poisson
 brackets (\ref{cor1}) and (\ref{cor2}) can be rederived from the "quantum" brackets
  associated with a dispersive counterpart of the Benney hierarchy.
  That means the loop correction can be viewed as the dispersive effect of the
  hydrodynamic type Poisson structure. Let us consider the "quantum"
   Lax operator of the form
\[
K=\ep\pa+u^1+(\ep\pa)^{-1}u^2,\qquad \pa\equiv \pa/\pa X
\]
which is just the Lax operator of the KB hierarchy discussed in \cite{KO,OS}
 under the scaling (\ref{scaling}).
 The bi-Hamiltonian structure associated with $K$ has been obtained by Oevel and Strampp
  \cite{OS} as follows
 \[
 \{I,J\}_i=\int \res\left[\frac{\de I}{\de K}\Omega_i\left(\frac{\de J}{\de K}\right)\right],
 \qquad i=1,2
 \]
where $I$ and $J$ are functionals of $K$ and the Hamiltonian maps $\Omega_i$ are defined by
 \bean
\Omega_1: \frac{\de I}{\de K}&\to& \left[\left(\frac{\de I}{\de K}\right)_{\ge 1}, K\right]-
 \left(\left[\frac{\de I}{\de K}, K\right]\right)_{\ge -1}\\
 \Omega_2: \frac{\de I}{\de K}&\to& \left(K\frac{\de I}{\de K}\right)_+K-
 K\left(\frac{\de I}{\de K}K\right)_+-\left[\left(K\frac{\de I}{\de K}\right)_0, K\right]-
 \left(\left[\frac{\de I}{\de K}, K\right]\right)_{-1}K\\
 & &+\left[\int^X\res\left[\frac{\de I}{\de K},K\right], K\right]
 \eean
 with
 \[
\frac{\de I}{\de K}=\frac{\de I}{\de u^2}+(\ep\pa)^{-1}\frac{\de I}{\de u^1}.
\]
 Using the Hamiltonian flows $\pa K/\pa t=\Th_i(\de H/\de K)$,
  we can easily read off the  "quantum" Poisson brackets
 \bean
  \{u^1(X),u^1(Y)\}_1&=&0,\\
  \{u^1(X),u^2(Y)\}_1&=&\de'(X-Y),\\
  \{u^2(X),u^1(Y)\}_1&=&\de'(X-Y),\\
  \{u^2(X),u^2(Y)\}_1&=&0
 \eean
 for the first structure and
 \bean
  \{u^1(X),u^1(Y)\}_2&=&2\de'(X-Y),\\
  \{u^1(X),u^2(Y)\}_2&=&u^1_{X}\de(X-Y)+u^1\de'(X-Y)+\ep\de''(X-Y),\\
  \{u^2(X),u^1(Y)\}_2&=&u^1\de'(X-Y)-\ep\de''(X-Y),\\
  \{u^2(X),u^2(Y)\}_2&=&u^2_{X}\de(X-Y)+2u^2\de'(X-Y)
 \eean
for the second. As a result, the first structure gets no correction, whereas the
 second structure receives a first order correction. So far, everything is exact.
 However if we define the following substitution for the flat coordinates $t^\alpha$:
 \bea
  t^1(T)&=&u^1- \ep(\ln u^2)_X-\frac{\ep^2}{24}\left(\frac{u^1_{X}}{u^2}\right)_X+
 \frac{\ep^3}{72}\left[\frac{(\ln{u^2})_{XX}u^2-(u^1_{X})^2}{(u^2)^2}\right]_X+O(\ep^4),\no\\
 t^2(T)&=&u^2-\frac{\ep}{2}  u^1_{X}+\frac{3 \ep^2}{8} (\ln u^2)_{XX}+\frac{11\ep^3}{144}
 \left(\frac{u^1_{X}}{u^2}\right)_{XX}
 +O(\ep^4)
 \label{exp}
 \eea
 where the right-hand side of $t^{\alpha}$ is constructed from $\pa^iv^{\alpha}/\pa
  (T^Q)^i|_{v^\alpha=u^\alpha}$,
then a straightforward but lengthy calculation shows that the first and the second
 Poisson brackets for $t^\alpha$ coincide with  equations (\ref{cor1}) and (\ref{cor2})
  modulo  $O(\ep^4)$.
   Furthermore, using (\ref{exp}), it is not hard to check that the  dispersion expansion of
     the Hamiltonians and hierarchy flows defined by the KB hierarchy and
     those defined by $H_{P,n}$ and $T^{P,n}$-flow coincide, modulo $O(\ep^4)$.
    In this sense, the parameter $\ep$ of genus
    expansion characterizes   the effect of dispersion.\\
  {\bf Remarks 5.\/} The free energy and $G$-function associated with the bi-Hamiltonian structure
  of Drinfeld-Sokolov reduction of Lie algebra $B_2$ are \cite{DZ}
  \[
 F=\frac{1}{2}(t^1)^2t^2+\frac{1}{15}(t^2)^5,\qquad
 G=-\frac{1}{48}\log t^2
  \]
where the $G$-function of $B_2$ is also of logrithmic type. However,
 in contrast with the Benney hierarchy, the second Poisson brackets of the dispersion expansion
coincide with those of DZ brackets only up to $\ep^0$ \cite{DZ}. This inconsistency
 also pointed out in \cite{EYY} by considering the commuting flows.

\section{Conclusions}
We have studied several interesting properties associated with the bi-Hamiltonian structure
 of the Benney hierarchy. Starting with the Poisson brackets of
hydrodynamic type we obtain the structure coefficients of an associative algebra
 characterizing the associated Frobenius manifold. This  implies that there exist a function
  which generates the structure coefficients and this function in fact
can be viewed as the genus zero primary free energy of a TFT.
 It turns out that the topological correlation functions at genus
  zero can be constructed by using the LG formulation.
  After appropriately define the  two-point correlation functions as the Benney variables,
   the genus zero topological recursion relation of the TFT turns out to be the Benney equations.
    Furthermore, we use the twistor construction to derive the string equation
     associated with the TFT, which describe the dynamics of correlation functions in
     the full phase space. Finally, based on the approach of Dubrovin and Zhang
     we obtain the genus one correction of the Poisson brackets. We  show that the same result
     can be reached by analyzing the OS brackets of
      the KB hierarchy in semi-classical limit.

Inspite of the results obtained, there are some interesting issues deserve more investigations.
\begin{itemize}
  \item One knows that there exists a Legendre-type transformation between the free
   energy (\ref{free}) and that of topological $CP^1$ model \cite{EHY}. In fact,
    one can check that the DZ brackets of the Benney hierarchy and those of dToda system in $CP^1$
    model \cite{DZ,EYY} coincide only up to $\ep^0$ via the Legendre-type transformation.
     A natural question is: Can we find a "quantized" version of the Legendre transformation
       between these two dispersionless hierarchies so that their DZ brackets
       match up to $g \geq 1$?
\item In the deviation of the string equation we borrow the
method of twistor construction to obtain the result. In\cite{DZ2},
 using recursion procedure,  Dubrovin and Zhang establish the
Virasoro constraints of genus zero for arbitrary Frobenius manifold.
 Then it would  be interesting to know whether the solution of string equation of the
  Benney hierarchy satisfies Virasoro constraints using  the methods developed
   in \cite{Kr1,TT}. Also, we wonder that whether there exists a matrix model
    associated with the Benney hierarchy such that the large-$N$
     limit of that reproduces the genus expansion  of the hierarchy flows.
 \item Recently, the genus-two free energy ${\mathcal F\/}_2$ of the Benney hierarchy
has been obtained in \cite{Du6} (see also \cite{EGX}) by combining genus-two topological
 recursion relations and Virasoro constraints \cite{EX}.
  So we should extend the expansions  (\ref{cor1})
  and (\ref{cor2}) to $O(\ep^6)$ when the genus-two correction is included.
   Then it is quite interesting to see whether the expansions (\ref{exp}) can be extended
    so that the OS brackets are matched with the DZ brackets
    up to  $O(\ep^6)$ after appropriate differential substitutions.
      \end{itemize}
We hope to report on these issues in a forthcoming paper.\\

{\bf Acknowledgments\/}\\
  J.H.C is thankful for Professor Dubrovin's stimulating conversations
 during the conference  "Integrable systems in Differential Geometry," held
 at Univ. of Tokyo, July 17-July 21, 2000. He also thanks for the support of the
  Academia Sinica. M.H.T thanks for the National Science Council of Taiwan under
   Grant No. NSC 89-2112-M194-018 for support.

\newpage

\appendix
\section{Proof of Theorem 1}

 {\it Proof.\/}
 Let us first derive the canonical Poisson relations (\ref{unit}). By the chain rule , we have
\begin{equation}
\left(
\begin{array}{cc}
\frac{\partial f(\mu, M)}{\partial \mu} & \frac{\partial f(\mu, M)}{\partial M} \\
 \frac{\partial g(\mu, M)}{\partial \mu} & \frac{\partial g(\mu, M)}{\partial M}
\end{array}
\right ) \left(
\begin{array}{cc}
\frac{\partial \mu }{\partial p} & \frac{\partial \mu }{\partial X}\\
 \frac{\partial M}{\partial p} & \frac{\partial M }{\partial X}
\end{array}
\right )
=
\left(
\begin{array}{cc}
\frac{\partial {\ti {f}}(\ti {\mu}, \ti {M})}{\partial \ti {\mu}} &
 \frac{\partial {\ti {f}}(\ti {\mu}, \ti {M})}{\partial \ti {M}} \\
  \frac{\partial \ti {g}(\ti {\mu}, \ti {M})}{\partial \ti {\mu}} &
   \frac{\partial \ti {g}(\ti {\mu}, \ti {M})}{\partial \ti {M}}
\end{array}
\right ) \left(
\begin{array}{cc}
\frac{\partial \ti {\mu} }{\partial p} & \frac{\partial \ti {\mu} }{\partial X}\\
 \frac{\partial \ti {M}}{\partial p} & \frac{\partial \ti {M} }{\partial X}
\end{array}
\right )  \label{det}
\end{equation}
Taking the determinant of both hand sides of this equation and using relations (\ref{pois}),
 we get
\begin{equation}
\{\mu, {M} \}=\{ \tilde{\mu}, \tilde{M} \}. \label{pois1}
\end{equation}

One can calculate the left hand side as
\begin{eqnarray*}
\{\mu, M \} &=& \frac{\partial \mu}{\partial p} \frac{\partial M}{\partial X}-
 \frac{\partial M}{\partial p} \frac{\partial \mu}{\partial X}, \\
 &=& \frac{\partial \mu}{\partial p} \left [\left (\frac{\partial
M}{\partial \mu} \right )_{w_{n}(T,\ti {T})\hspace{2mm} \mbox{fixed}}
 \frac{\partial \mu}{\partial X}+1+\sum_{i=1}^{\infty}
\frac{\partial w_{i}(T,\ti {T})}{\partial X} \mu^{-i} \right] \\
 &-& \frac{\partial \mu}{\partial X}  \left (\frac{\partial M}
 {\partial \mu} \right )_{w_{n}(T, \ti {T})\hspace{2mm} \mbox{fixed}}
\frac{\partial \mu}{\partial
 p},\\
&=&1+(\mbox{negative powers of} \,\  p)
\end{eqnarray*}
where we have used the fact that the terms containing $\left (\frac{\partial M}{\partial \mu}
 \right )_{w_{n}(T, \ti {T})\hspace{2mm} \mbox{fixed}} $ in
  the last line cancel. Similar calculations can show that the right hand side contains only the
non-negative powers of $p$. Therefore the both hands of (\ref{pois1}) are equal to $1$,
 which proves the canonical relations (\ref{unit}).\\
  \indent The Lax  equations with respect to $T_n$ are proved as follows.
   Differentiating equations (\ref{fun}) by $T_n$
gives
\begin{equation}
\left(
\begin{array}{cc}
\frac{\partial f(\mu, M)}{\partial \mu} & \frac{\partial f(\mu, M)}{\partial M} \\
 \frac{\partial g(\mu, M)}{\partial \mu} &
\frac{\partial g(\mu, M)}{\partial M}
\end{array}
\right ) \left(
\begin{array}{c}
\frac{\partial \mu }{\partial T_n} \\ \frac{\partial M}{\partial T_n}
\end{array}
\right )
=
\left(
\begin{array}{cc}
\frac{\partial {\ti {f}}(\ti {\mu}, \ti {M})}{\partial \ti {\mu}} &
 \frac{\partial {\ti {f}}(\ti {\mu}
 , \ti {M})}{\partial \ti {M}} \\ \frac{\partial \ti {g}(\ti {\mu},
  \ti {M})}{\partial \ti {\mu}} & \frac{\partial
   \ti {g}(\ti {\mu}, \ti {\mathcal} M)}{\partial \ti{M}}
\end{array}
\right ) \left(
\begin{array}{c}
\frac{\partial \ti {\mu} }{\partial T_n} \\ \frac{\partial \ti {M}}{\partial T_n}
\end{array}
\right )  \label{flow}
\end{equation}
Using (\ref{det}), we can rewrite (\ref{flow}) as
 \bea
 \left(
\begin{array}{cc}
\frac{\partial \mu }{\partial p} & \frac{\partial \mu }{\partial X}\\
 \frac{\partial M} {\partial p} & \frac{\partial  M}{\partial X}
\end{array}
\right )^{-1} \left(
\begin{array}{c}
\frac{\partial  {\mu} }{\partial T_n} \\ \frac{\partial  M}{\partial T_n}
\end{array}
\right)
=
\left(
\begin{array}{cc}
\frac{\partial \ti {\mu} }{\partial p} & \frac{\partial \ti {\mu} }{\partial X}\\
 \frac{\partial \ti {M}}{\partial p} & \frac{\partial \ti {M} }{\partial X}
\end{array}
\right )^{-1} \left(
\begin{array}{c}
\frac{\partial \ti {\mu} }{\partial T_n} \\ \frac{\partial \ti {M}}{\partial T_n}
\end{array}
\right ) \label{det1}
 \eea
  Since the the determinants of the $2 \times 2$
 matrices on both sides are $1$, the inverse can also be written explicitly.
  In components, thus, the above matrix (\ref{det1}) relation gives
 \begin{eqnarray}
 \frac{\partial M}{\partial X} \frac{\partial
 \mu}{\partial T_n} -\frac{\partial \mu}{\partial X} \frac{\partial
 M}{\partial T_n} &=& \frac{\partial \tilde
 {M}}{\partial X} \frac{\partial \tilde {\mu}}{\partial
 T_n} -\frac{\partial \tilde {\mu}}{\partial X} \frac{\partial \tilde
 {M}}{\partial T_n}, \label{cal1} \\
 \frac{\partial
 M}{\partial p} \frac{\partial \mu}{\partial T_n}
 -\frac{\partial \mu}{\partial p} \frac{\partial
 M}{\partial T_n} &=& \frac{\partial \tilde {
 M}}{\partial p} \frac{\partial \tilde {\mu}}{\partial T_n}
 -\frac{\partial \tilde {\mu}}{\partial p} \frac{\partial \tilde
 {M}}{\partial T_n}.
 \label{cal2}
 \end{eqnarray}
The left hand sides of equations (\ref{cal1}) and (\ref{cal2})
  can be calculated just as we have done above for derivatives
in $(p, X)$. For the  equation of (\ref{cal1}),
\begin{eqnarray}
& &\frac{\partial M}{\partial X} \frac{\partial \mu}{\partial T_n}
 -\frac{\partial \mu}{\partial X} \frac{\partial
M}{\partial T_n}  \no \\
 &=& \left [\left (\frac{\partial M}{\partial \mu} \right )_{w_{i}(T, \ti {T})\hspace{2mm}
  \mbox{fixed}}
\frac{\partial \mu}{\partial X}+1+\sum_{i=1}^{\infty}
 \frac{\partial w_{i}(T, \ti {T})}{\partial X} \mu^{-i} \right]
\frac{\partial \mu}{\partial T_n}  \no
 \\ &-& \frac{\partial \mu}{\partial X}
 \left [\left (\frac{\partial M}{\partial \mu}
\right )_{w_{i}(T, \ti {T})\hspace{2mm} \mbox{fixed}} \frac{\partial \mu}{\partial T_n}+
 n \mu^{n-1} +\sum_{i=1}^{\infty}
\frac{\partial w_{i}(T, \ti {T})}{\partial T_n} \mu^{-i} \right], \no
\end{eqnarray}
and terms containing $\left (\frac{\partial M}{\partial \mu} \right )_{w_{i}
 (T, \ti {T})\hspace{2mm} \mbox{fixed}} $ cancel.
Thus,
\begin{equation}
\frac{\partial M}{\partial X} \frac{\partial \mu}{\partial T_n} -\frac{\partial \mu}{\partial X}
 \frac{\partial M}{\partial T_n} = -\frac{\partial (\mu^n)_{\geq 1}}{\partial X}
  +(\mbox{powers of} \,\  p \leq 0). \label{rest1}
\end{equation}
Similar calculations show that the right hand side of (\ref{cal1})
 contains only positive powers of $p$. Therefore only powers
of $p \geq 1$ should survive. Hence
\begin{equation}
\frac{\partial M}{\partial X} \frac{\partial \mu}{\partial T_n} -\frac{\partial \mu} {\partial X}
 \frac{\partial M}{\partial T_n} = -\frac{\partial (\mu^n)_{\geq 1}}{\partial X}=
 -\frac{\partial  B_n}{\partial X}. \label{fina1}
\end{equation}
For the equation  (\ref{cal2}), we have similarly
\begin{eqnarray*}
\frac{\partial M}{\partial p} \frac{\partial \mu}{\partial T_n} -\frac{\partial \mu}
 {\partial p} \frac{\partial  M}{\partial T_n} &=& -\frac{\partial (\mu^n)_{\geq 0}}{\partial p}
  +(\mbox{negative powers of} \,\  p ), \\ &=& -\frac{\partial
(\mu^n)_{\geq 1}}{\partial p} +(\mbox{negative powers of} \,\  p ) .
\end{eqnarray*}
Similarly, noticing the partial derivative $\frac{\pa}{\pa p}$, we can also show
 that the right hand of (\ref{cal2}) has Laurent
expansion with only non-negative powers of $p$.
  Hence only nonnegative powers of $p$ should survive. Thus
\begin{equation}
\frac{\partial M}{\partial p} \frac{\partial \mu}{\partial T_n}
 -\frac{\partial \mu}{\partial p} \frac{\partial M}{\partial T_n} =
  -\frac{\partial (\mu^n)_{\geq 1}}{\partial p}=-\frac{\partial B_n}{\partial p}. \label{fina2}
\end{equation}
Using
\[ \{\mu, {M} \}=\{ \tilde {\mu}, \tilde{M} \}=1, \]
the equations (\ref{fina1}) and (\ref{fina2}) can be readily solved:
\begin{eqnarray*}
\frac{\partial \mu}{\partial T_n} &= & -\frac{\partial \mu}{\partial p}\frac{\partial B_n}
 {\partial X} + \frac{\partial \mu}{\partial X}\frac{\partial B_n}{\partial p}=\{B_n, \mu \}, \\
  \frac{\partial M}{\partial T_n} &= & -\frac{\partial M}{\partial p}\frac{\partial B_n}
  {\partial X} + \frac{\partial M}{\partial X}\frac{\partial B_n}{\partial p}=\{B_n, M \},
\end{eqnarray*}
which is nothing but $T_n$-flow part of the Lax equations (\ref{bar1}) and (\ref {bar2}).
 The $\ti {T}_n$-flow part of the Lax equations can be proved in the similar way.
  This completes the proof of the theorem.

\section{A deviation of $G$-fucntion}

For the quasi-homogeneous primary free energy (\ref{free}), one can introduce
 the canonical coordinates $u^1, u^2$ determined by
\[
\det(g^{ij}(t)-u\eta^{ij}(t))=0
\]
which gives
\[
u^1=t^1-2\sqrt{t^2},\qquad u^2=t^1+2\sqrt{t^2}.
\]
Then the topological metric $\eta^{ij}$ and its inverse $\eta_{ij}$ in the
 canonical coordinates are given by
 \be
\eta^{ij}(u)=\frac{\pa u^i}{\pa t^k}\frac{\pa u^j}{\pa t^l}\eta^{kl}(t)=
 \left(
  \ba{cc}
   -\frac{8}{u^2-u^1} & 0 \\
   0 & \frac{8}{u^2-u^1}
    \ea
 \right),\qquad
\eta_{ij}(u)= \left(
  \ba{cc}
   -\frac{u^2-u^1}{8} & 0 \\
   0 & \frac{u^2-u^1}{8}
    \ea
 \right).
\label{umetric}
 \ee
 It has been shown \cite{DZ} that the $G$-function can be expressed
 in the following formula
\[
G(t^1,t^2)=\log\frac{\tau_I}{J^{1/24}}
\]
where $J$ is the Jacobian of the transformation from the canonical coordinates to the flat ones,
 which is easily obtained as
\[
J=\det(\frac{\pa t^i}{\pa u^j})=\frac{1}{2}\sqrt{t^2}
\]
and $\tau_I$ is the tau-function of a solution in the theory of isomonodromy
 deformation defined by the quadrature \cite{JMMS}
 \be
d\log\tau_I=\sum_{i=1}^2H_idu^i \label{deftau}
 \ee
 where the quadratic Hamiltonian $H_i$ has the form
 \[
 H_i=\frac{1}{2}\sum_{j\neq i}\frac{V_{ij}^2}{u^i-u^j}
 \]
 with
 \[
 V_{ij}=-(u^i-u^j)\ga_{ij}(u),\qquad
 \ga_{ij}(u)=\frac{\pa_j\sqrt{\eta_{ii}(u)}}{\sqrt{\eta_{jj}(u)}}.
 \]
 Note that $\pa_i\equiv \frac{\pa}{\pa u^i}$ and $V_{ij}=-V_{ji}$.

From (\ref{umetric}) we have
 \[
  \ga_{ij}(u)=
   \left(
  \ba{cc}
    \frac{1}{2(u^1-u^2)} & \frac{-i}{2(u^2-u^1)} \\
    \frac{-i}{2(u^2-u^1)} & \frac{1}{2(u^2-u^1)}
    \ea
 \right),\qquad
 V_{ij}=
\left(
  \ba{cc}
    0 & -\frac{i}{2} \\
    \frac{i}{2} & 0
    \ea
 \right)
 \]
 which together with (\ref{deftau}) implies
 \[
\tau_I=(u^2-u^1)^{-1/8},
 \]
and hence, up to an additive constant
\[
G(t^1,t^2)=\log\frac{(u^2-u^1)^{-1/8}}{[(u^2-u^1)/8]^{1/24}}
 = -\frac{1}{12}\log t^2.
\]

\end{document}